\documentclass[12pt,paper]{emulateapj}

\usepackage{graphicx}
\usepackage{amsmath}
\usepackage{amssymb}
\usepackage{graphics}
\usepackage{graphicx}
\usepackage{ulem}

\def\MgII{\ion{Mg}{2}}
\def\HI{\ion{H}{1}}

\def\textbftemp{}
\def\textbfnew{}

\begin{document}

\title{The nature of damped Lyman-$\alpha$
and MgII absorbers\\ explored with their dust contents}

\author{ Masataka Fukugita$^{1,2}$ and Brice M\'enard$^{3,1,4}$ }

\affil{$^1$ {Kavli Institute for the Physics and Mathematics of the Universe, University of Tokyo, Kashiwa 277-8583, Japan}}
\affil{$^2$ { Institute for Advanced Study, Princeton, NJ 08540, USA} }
\affil{$^3$ { Department of Physics \& Astronomy, Johns Hopkins University, 3400 N. Charles Street, Baltimore, MD 21218, USA}}
\altaffiltext{4}{Alfred P. Sloan Fellow}

\begin{abstract}
  We estimate the abundance of dust in damped Lyman-$\alpha$ absorbers
  (DLAs) by statistically measuring the excess reddening they induce
  on their background quasars. We detect systematic reddening behind
  DLAs consistent with the SMC type reddening curve and inconsistent
  with the Milky Way type.  \textbftemp{We find that the derived
    dust-to-gas ratio is, on average, inversely proportional to the
    column density of neutral hydrogen, implying that the amount of
    dust is constant, irrespective of the column density of hydrogen.}
 \textbftemp{It means that}
 \textbftemp{the average metallicity is
inversely proportional to the column density of hydrogen,}
\textbfnew{unless the average dust-to-metal ratio varies with the hydrogen column density.}
  This indicates that the prime origin of 
  metals seen in \textbftemp{DLAs}
  is not by {\it in~situ} star formation, with which $Z\sim N_{\rm
    HI}^{+0.4}$ is expected from the empirical star formation law,
  contrary to our observation. We interpret the metals observed in
  absorbers being deposited dominantly from nearby galaxies by
  galactic winds ubiquitous in intergalactic space. 
  When extrapolating the relation between 
  \textbftemp{dust-to-gas ratio and \HI\ column density to lower column density, 
  we find a \textbftemp{value} which is consistent with what is observed 
  for \MgII\ absorbers.}
\end{abstract}


\section{Introduction}
\label{sec:introduction}

Whether intervening absorbers seen in quasar spectra are aggregates of
primordial material or results of the activity in galaxies is an
elementary problem. In a previous publication (M\'enard \& Fukugita
2012; hereinafter MF12), it was advocated that Mg II clouds are likely
to be a product of the activity of nearby galaxies with gas exported
by galactic winds.  This inference is based on the fact that the
observed dust abundance of \MgII\ clouds relative to gas takes a value
typical of galactic disks, while the star formation activity is not
observed nor expected in such clouds.

MF12 estimates that the global \HI\ abundance in \MgII\ clouds is
$\Omega_{\rm HI}({\rm MgII})\approx 1.5\times 10^{-4}$, which is
approximately 3\% of the fuel consumed by star formation by the
present epoch, or roughly 6\% at $z\approx 2$.  The amount of matter
expelled through galactic winds in actively star forming galaxies
\textbftemp{is often inferred to be comparable to} the star formation rate
(e.g., Heckman et al. 2000; Veilleux et al. 2005; Weiner et al. 2009).
The fraction of \HI\ in \MgII\ absorbers, which is 10 -- 20 \% of the
gas re-shed by stars as a whole, is not an unreasonable amount as a
product resulting from the star formation activity in galaxies. It is
also shown that dust in \MgII\ clouds accounts for half the amount of
dust estimated to reside outside galaxies.

The analysis gives the example that the dust abundance, as explored by
extinction of light rays passing through the absorbers, provides us
with a useful indicator of the heavy element abundance, assuming that
photometry is accurate. This suggests that more could be learned
from dust studies \textbftemp{of} other classes of absorbers.

In this paper we focus on damped Lyman-$\alpha$ absorbers (DLAs). 
The global mass density of \HI\ in DLA clouds
has been estimated to be $\Omega_{\rm HI}({\rm DLA})\approx
(4-10)\times 10^{-4}$ at $z\approx 2$ by Prochaska \& Wolfe (2009)
using the quasar spectroscopic data of SDSS DR5, and more recently by
Noterdaeme et al. (2012) using SDSS DR9.  This quoted range 
\textbftemp{
arises from different treatments of the continuum around the damped Lyman
$\alpha$ line, as well as survey path length and completeness.
}
Whichever value is taken, the DLA mass density is
significantly larger than the \HI\ mass density in the \MgII\
absorbers, which is estimated to be $\Omega_{\rm HI}({\rm
  MgII})\approx 1.5\times 10^{-4}$ (MF12). 
\textbftemp{\HI\ mass density is clearly larger than that can be associated with
stellar activity.}

Studies of individual DLAs have indicated their metallicity [Fe/H] to
be in the range $-0.5$ to $-2$ (e.g., Prochaska et al. 2003; Rafelski
et al. 2012), an order of magnitude lower than that of galaxies. This
contrasts \textbftemp{with} \MgII\ absorbers, which MF12 showed to have metallicity
of the order of solar.  These observations may be taken in favour of
the interpretation that there would be two distinct populations of
absorbers as to their origin: DLAs belonging to one
and \MgII\ absorbers to another. This also induces the question as to where
the metallicity of DLAs arises from.

There have been a few attempts to detect reddening of quasars behind
DLAs (\textbftemp{Fall et al. 1989, Pei et al. 1991, Murphy \& Liske 2004,} Vladilo et al. 2008; Frank \& P\'eroux 2010, Khare et al. 2011).
The results have not always led to positive detections. The difficulty
is that the reddening signal is small, of the order of a few
hundredths of magnitude, whereas sample variation due to objects is an
order of magnitude larger. For photometric studies, one needs to
accurately define \textbftemp{the mean colour of reference quasars}.
With spectroscopic work one needs
accurate sensitivity calibrations over \textbftemp{a wide wavelength range}.

In this paper we estimate the mean reddening effects induced by DLAs
by comparing broad band flux of the quasar light showing Lyman
$\alpha$ absorptions with damped wings to that without absorbers. We
use accurately calibrated SDSS broad band photometry, and we limit
the redshift range to minimise the scatter of fiducial
quasar colours.  We are concerned with a photometric accuracy 
smaller than 0.1 mag. It turns out that choosing the right range of
\textbftemp{both quasar and absorber redshifts} is important in keeping
the errors of colours small. Specifically, care must be made so that
passbands are away from the Lyman edge or the Lyman $\alpha$ line for
absorbers.
\textbftemp{We study whether the dust-to-gas ratio of DLAs is correlated
  with their hydrogen column density, and also examine if any
  difference is seen between \MgII\ absorbers and DLAs in the
  metallicity \HI\ column density relation.}

\section{The data}

\begin{figure}[t]
\includegraphics[width=.48\textwidth]{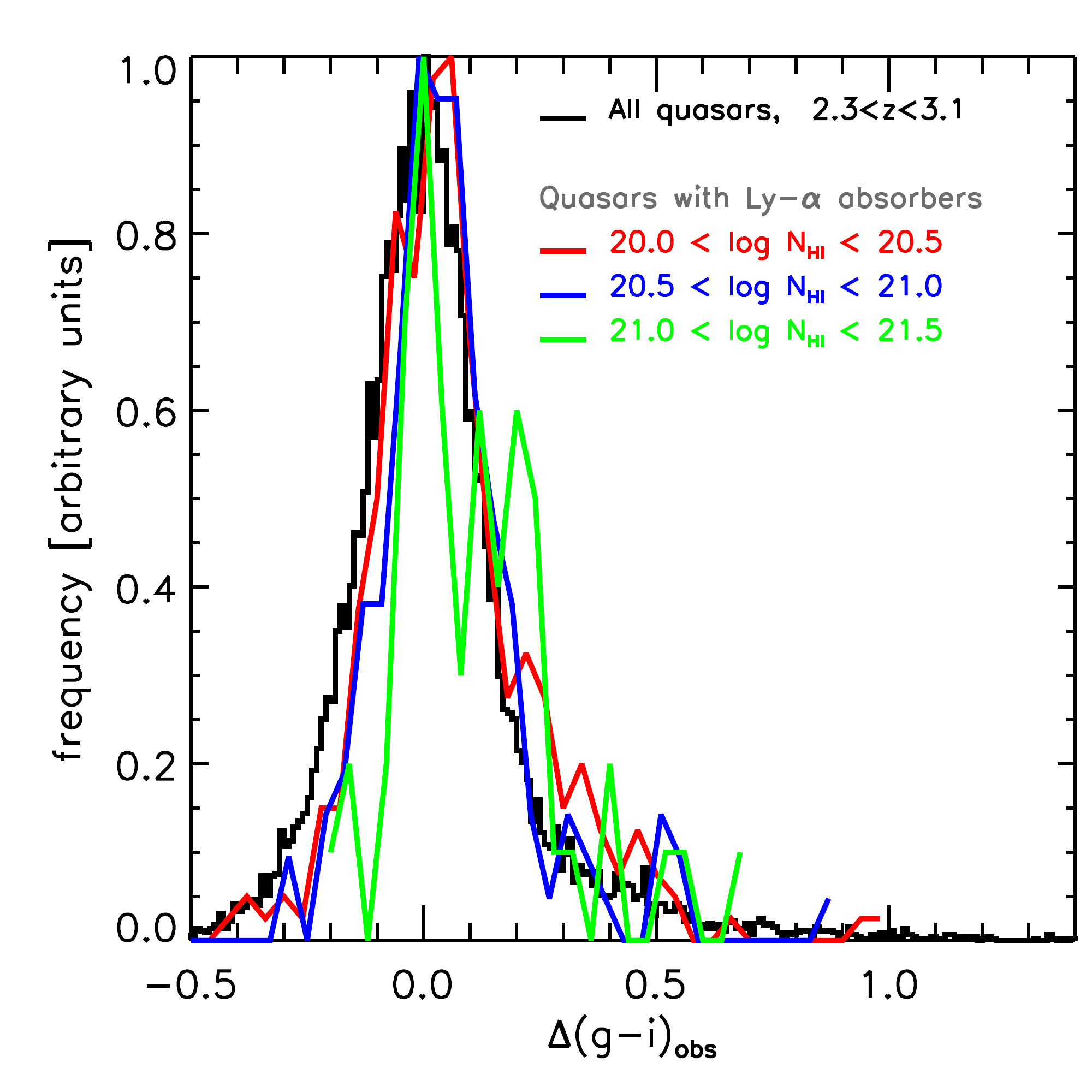}
\caption{Distribution of colour excess $\Delta (g-i)$ for quasars
  showing Lyman $\alpha$ absoption for three bins of hydrogen column
  densities.  The thin (black) histogram shows the distribution for
  the quasar sample with $2.3<z<3.1$.  }
\label{fig:1}
\end{figure}

We use the quasar catalogue compiled by P{\^a}ris et al. (2012) based
on the ninth data release of the Sloan Digital Sky Survey III
(hereinafter DR9, Ahn et al. 2012).  Noterdaeme et al. (2012) analyzed
these 87,822 quasars and identified 12,081 clouds with \HI\ column
density $N_{\rm HI}\geq 10^{20}{\rm cm^{-2}}$ for redshift $z\geq
1.9$.  The column density is estimated using Voigt profile fitting.
About 57\% of the clouds (6839) have $N_{\rm HI}\geq10^{20.3} {\rm
  cm^{-2}}$ with damped wings. The quasar sample
consists of a union of quasars arisen from different 
selections. Since accurate photometry and accurate colour property are
of primary importance for our analysis, \textbftemp{ we select objects
  with the flag {\tt UNIFORM=1} which produces a homogeneously selected sample of quasars (see section 4.3 of P\^aris et al. 2012).
}
This uniform sample contains 23,499 quasars.

\textbftemp{
Since we want to detect very small values of reddening, it is of
crucial importance to define accurately reference quasar colours.  
}
\textbftemp{To do so we first inspect the distribution of quasar colours as a function of redshift and observe that
the scatter increases substantially at $z>3.1$. We therefore restrict our analysis to quasars selected in the redshift range of quasars to $2.3<z<3.1$ where the lower limit is motivated by absorber selection, as explained below. In this range,}
we find that the difference between the mean and median color as a function of redshift
is smaller than 
\textbftemp{
about 0.01
mag for $z<3.1$. Using higher-redshift quasars increases the scatter in colours. We therefore do not include them in our analysis.
}

Zhu \& M\'enard (2012) identified in the DR9 uniform quasar sample
10,877 \MgII\ absorbers with rest equivalent widths $W_0>0.3$ \AA, of
which 6,635 have $W_0>0.8$ \AA, in the redshift range $z=0.36$ and 2.5.
\textbftemp{
We note that no velocity cuts are applied to the \MgII\ absorbers.
}
We restrict our analysis to the \MgII\ absorbers with $W_0>0.8$ \AA,
below which the sample completeness drops to $\lesssim 50$\% (Nestor
et al. 2005), although completeness is not important to our
analysis.

The DLA and \MgII\ samples overlap in the redshift range $1.9\leq z\leq 2.5$. We restrict primarily our analysis to the range $2.1\leq
z_{\rm abs}\leq 2.3$, avoiding redshift tails of both DLA and \MgII\
absorber distributions. With this selection we can minimise the
effect of Lyman opacity, which contributes to \textbftemp{the} colours we choose. This
overlap allows us to study the relation between the two populations of
absorbers.  This redshift interval contains in the common quasar
sample 1211 DLAs and 680 Mg II absorbers, and 150 among those are
common in both samples.  These statistics indicate that 12\% of DLAs
show \MgII\ absorption lines detected with $W_0>0.8$ \AA, and
conversely 22\% of Mg II absorbers are DLAs.

\section{Dust reddening and metallicity of DLA}

As was done for \MgII\ absorbers in MF12,
we measure mean reddening of background quasars
induced by intervening clouds. 
We estimate the colour excess
\begin{equation} 
\Delta(g-i)_{\rm obs}= (g-i) - \langle (g-i) \rangle\;,
\label{eq:gi_def}
\end{equation}
where the expectation value is computed using a median
to avoid largely scattered data points that occasionally 
happen
\footnote{\textbftemp{In MF12 the reference quasar sample is constructed
    from quasars without detectable absorption lines. In the present
    work the reference sample includes all quasars. Since the quasar
    sample is dominated by quasars without absorbers, the difference
    between the two selections has negligible effects on the mean
    reference quasar colour. We find it to differ by less than
    $<3\times 10^{-3}$ mag, which is undetectable in our analysis. We
    also note that the quantity $\Delta(g-i)$ given in the DR9 quasar
    catalogue (P\^aris et al. 2012) is obtained by taking a mode of
    the distribution. This differs from our definition
    (Eq.\ref{eq:gi_def}) which uses a median estimate. The difference
    between the two estimators is of order 0.05 mag.}  },
rather than the mean.

Dividing the quasar sample into a $\delta z=0.25$ bin, we set the
median of $g-i$ to zero at each bin. \textbftemp{As a first test we} compare
the colours of all quasars with those that do not show DLA signatures.
Here, the effect of reddening due to absorbers is diluted by the
presence of a larger population that does not show DLA signatures by
about 10 times.  The difference is smaller than 0.02 mag, \textbftemp{which is
our current goal in view of the accuracy of photometry.}  This test,
which shows null detection below the noise level and the homogeneity
of colour, \textbftemp{reassures us that we can detect a colour change if it is
larger than 0.02 mag}.

In Figure 1 we show the distributions of $\Delta(g-i)$ for our quasars
with and without hydrogen absorbers.  \textbftemp{ Here and hereafter we
  take the sample down to $N_{\rm HI}=10^{20}{\rm cm}^{-2}$ to include
  sub-DLA clouds to examine if there is any change in the trend across
  the DLA threshold\footnote{Where
    we referred to DLA, the sample is restricted to $N_{\rm
      HI}\ge10^{20.3}{\rm cm}^{-2}$.} $N_{\rm HI}=10^{20.3}{\rm cm}^{-2}$
  }.  
The distributions with hydrogen
absorbers for three bins of $N_{\rm HI}$ overlap with each other.
When compared to the reference $\Delta(g-i)$ distribution of quasars
(thinner black curve), we see that those with Lyman absorption are
shifted redwards by an amount of about 0.05 mag.  The sample
dispersion of $\Delta(g-i)$ colour is 0.2 mag which indicates that
one may detect a reddening signal, say, as small as 0.02 mag only when
averaged over 100 or more absorbers.  Each measurement, which
sometimes gives negative values, does not tell about reddening for
individual quasars.  Their colour shift are 0.05 mag or even less,
which is buried in much larger scatters, arising from both variation
among individual quasars and measurement errors. Averaging over a
large sample is needed to detect a meaningful reddening signal.

\begin{figure}[t]
\includegraphics[width=.48\textwidth]{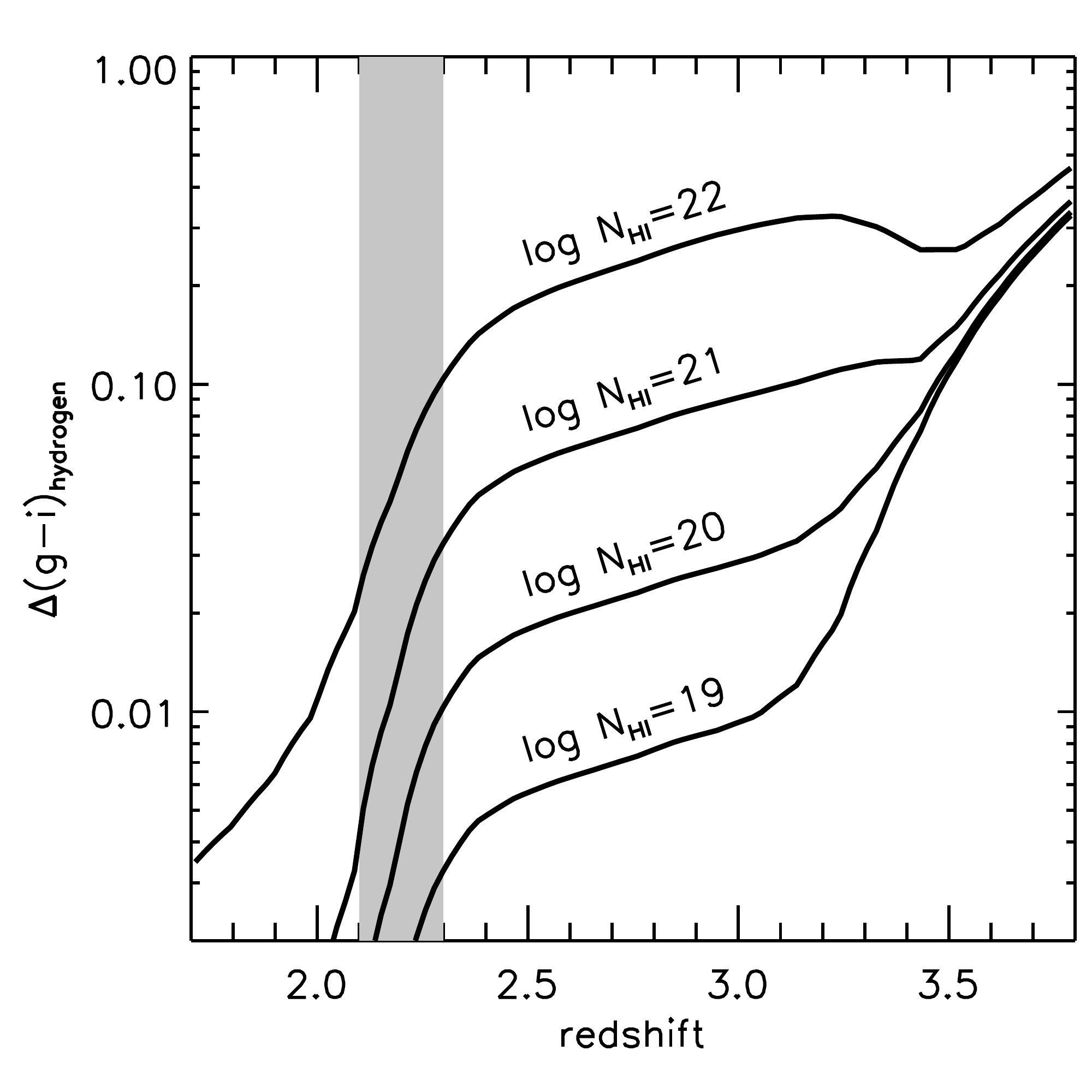}
\caption{Expected colour excess $\Delta (g-i)$ due to Lyman opacity
  in absorbers 
as a function of redshift for specified values of $N_{\rm HI}$. }
\label{fig:2}
\end{figure}

\begin{figure}[t]
\includegraphics[width=.5\textwidth]{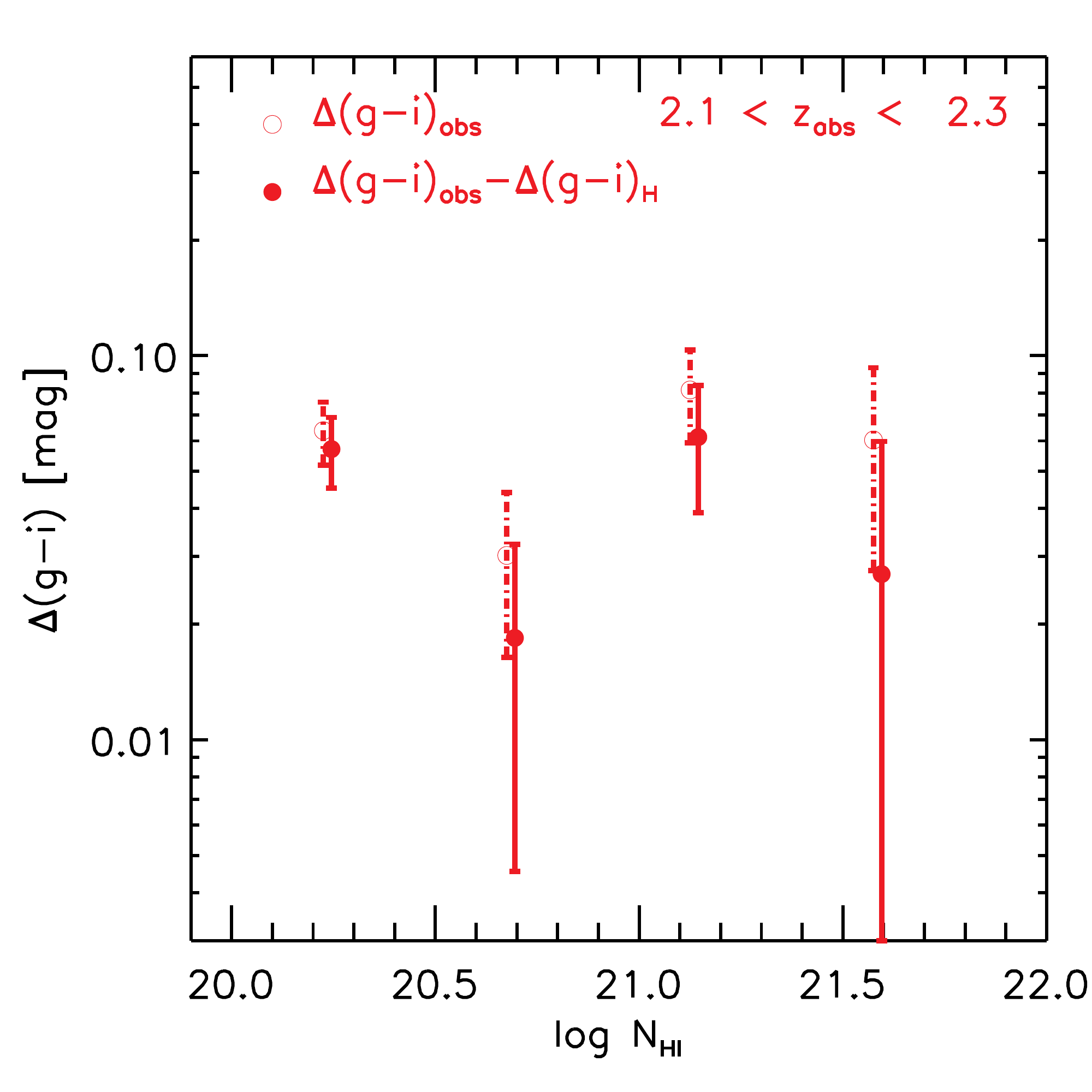}
\caption{Colour excess $\Delta (g-i)$ caused by Lyman opacities in
  absorbers with $2.1<z<2.3$ as a function of $N_{\rm HI}$. The open circles show the
  measured median and  the filled
  circles show the values corrected for hydrogen opacity.}
\label{fig:3}
\end{figure}

\textbftemp{We expect a} colour excess mainly caused by \textbftemp{both} dust reddening and by hydrogen Lyman opacity. Figure 2 shows the expected 
$g-i$ reddening due to Lyman opacity as a function of redshift for specified values of
$N_{\rm HI}$.  One sees the first rise due to the Lyman $\alpha$
excitation and the second due to the Lyman edge.  When the Lyman edge
is met, $\Delta(g-i)$ due to hydrogen ionisation is larger than that
from dust, which is of the order of 0.05 mag (as we see in Figure 3
below). Even away from the Lyman edge, reddening due to the Lyman
excitation is non-negligible and must be taken into account.  
\textbftemp{
  We define reddening due to dust as
\begin{equation} 
\Delta(g-i)_{\rm dust}= \Delta(g-i)_{\rm obs} - \Delta(g-i)_{\rm H}\;.
\label{eq:gi_dust}
\end{equation}
}
\textbftemp{We note that restricting our analysis to $z\leq2.3$ has the 
advantage of significantly reducing} the effect of Lyman $\alpha$ opacity in the
dust reddening estimate.  Figure 3 shows measured reddening
$\Delta(g-i)$ as a function of \textbftemp{hydrogen column density} for
absorbers $2.1<z_{\rm abs}<2.3$ for both \textbftemp{ raw measured value
  (open circles) and value after hydrogen opacity subtracted (filled
  circles).  }  As expected from Figure 1 the dependence of
$\Delta(g-i)$ with hydrogen column \textbftemp{density stays approximately
  constant at $\approx$0.05 mag.}  \textbftemp{The colour excess due to
  dust reddening $\Delta(g-i)_{\rm dust}$ is of order of a few percent
  and corresponds to a rest-frame reddening value of about
  $E(B-V)\approx 0.01$, assuming a SMC extinction curve}.  This is in
agreement \textbftemp{with Vladilo et al. (2008)}.  
\textbftemp{We also remark that
this level of reddening is consistent with the value reported by
Frank \& P\'eroux (2010) when they restrict their analysis to $z<2.2$.}

\textbfnew{
Broad-band reddening effects due to the presence of metal lines was studied by M\'enard \&
Fukugita (2012) using both calculations and measurements. In these studies we showed
that the combination of MgII and FeII lines (some of the strongest absorption features
in the UV) induce magnitude changes at a level below one percent and we were able to detect
this effect using strong MgII absorbers spanning the range $0.5<z<2$.
Pieri et al. (2014) presented composite spectra of Lyman-$\alpha$ absorbers. Their results show
that the metal lines that would fall in the $g$ and $i$ bands would be weaker than the combination
of MgII and FeII lines. The presence of metal lines is therefore not expected to appreciably
affect the measured color excess of quasars behind absorbers in the current analysis.
}

\begin{figure*}[t]
\begin{center}
\vspace{-1.5cm}
\includegraphics[width=\textwidth,angle=0]{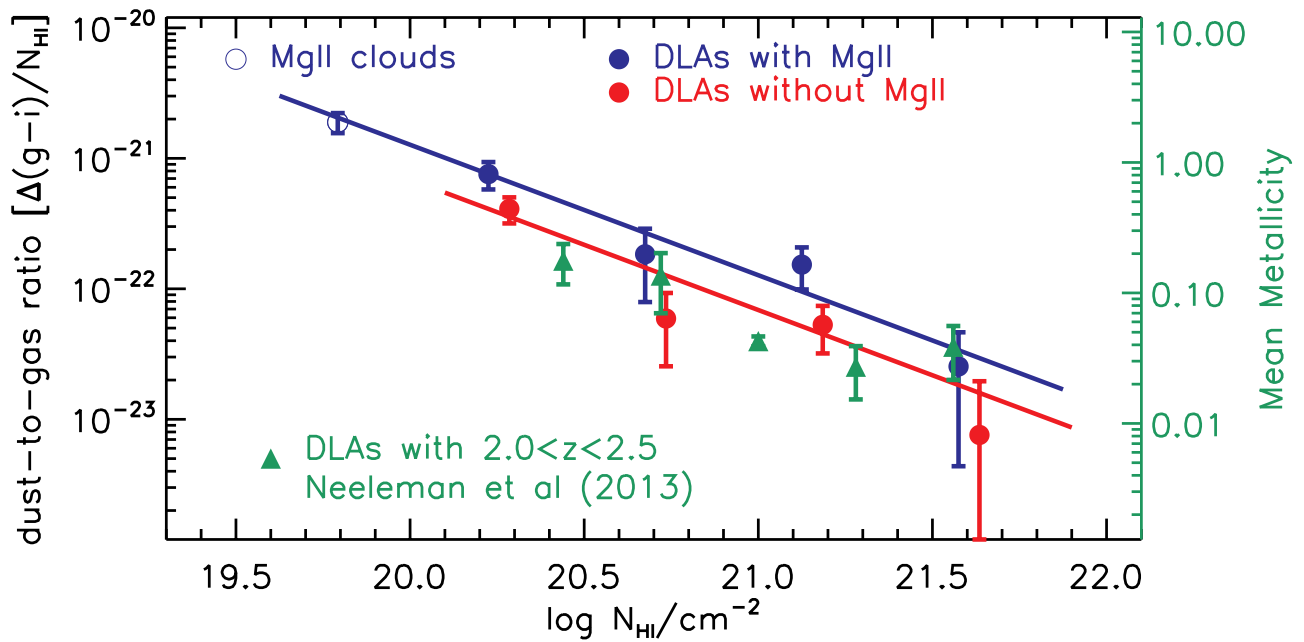}
\vspace{-2.5cm}
\caption{Median dust-to-gas ratios for the absorbers as a function of
  $N_{HI}$ for systems with $2.1<z<2.3$. Solid blue circles are
  median values for clouds showing the \MgII\ absorption with
  0.8\AA$<W_0$, solid red circles are that for clouds without MgII,
both estimated from median reddening in $g-i$ colour.
  The lines are fit with the slope fixed to $Z\propto N_{HI}^{-1}$. 
 The open symbol at $\log N_{HI}=19.8$ is for MgII
  absorbers. The right axis shows the inferred metallicity according to Eq.~5. 
  Green data symbols show spectroscopic estimates for
  metallicity for DLA in $2.1<z<2.3$. taken from Neeleman et al. (2013)}
\label{fig:}
\end{center}
\end{figure*}

\textbftemp{Having both reddening and Lyman-$\alpha$ absorption
  measurements we estimate the average dust-to-gas ratio of the
  systems as measured by ${\Delta(g-i)_{\rm dust}}/{N_{\rm HI}}$.
  Figure 4 shows this ratio as a function of $N_{\rm HI}$ for
  absorbers at $2.1\leq z_{\rm abs}\leq2.3$.  }
Circles are median estimates and the error bars are \textbftemp{computed} from bootstrap resampling.
We show the dust-to-gas ratio separately for clouds
showing \MgII\ features and those that do not.
We observe that the two trends are parallel.
The median is \textbftemp{well} represented by
\begin{equation}
\frac{\Delta(g-i)_{\rm dust}}{N_{\rm HI}}
\propto N_{\rm HI}^{-1.0\pm 0.2},
\end{equation}
for DLAs showing \MgII\ features, and
\begin{equation}
\frac{\Delta(g-i)_{\rm dust}}{N_{\rm HI}}
\propto N_{\rm HI}^{-1.3\pm 0.3},
\end{equation}
for DLAs without \MgII\ features, both indicating ${\Delta(g-i)_{\rm
    dust}}/{N_{\rm HI}} \propto 1/N$.  This means that the total metal
abundance along the column hardly depends on the hydrogen column
density of the cloud, as we noted in Figures 1 and 3 above.  For this
redshift range the contribution from Lyman alpha opacity is small, as
mentioned, which is anyway subtracted out (if not the power appears
more like $N_{\rm HI}^{-0.8}$).  We do not see any break beyond the
error in this
correlation for the range $N_{\rm HI}=10^{20.0}-10^{21.8} {\rm cm}^{-2}$.

We note that the two curves in Figure 4, one with \MgII\ signature 
and the other
without, obey parallel 
relations, meaning that all
DLAs are dusty, irrespective of whether they show significant \MgII\
absorption or not. Those that show \MgII\ absorption have somewhat
larger metallicity, approximately by a factor of 3. Whether a system
shows \MgII\ absorption depends solely on the total \MgII\ abundance
in the relevant column. When the total abundance of \MgII\ exceeds
some threshold (roughly $N_{\rm Mg}\simeq3\times 10^{15}{\rm
  cm}^{-2}$), the clouds are identified as \MgII\ absorbers with
$W_0>0.8\,$\AA.

One data point added below our analysis threshold $\log N_{\rm HI}<20$
refers to \MgII\ clouds in the $W_0>0.8$\AA~ sample in
MF12. The \HI\ column density is not directly measured for these
absorbers. We infer it invoking the $W_0$ - $N_{\rm H}$ relation
(M\'enard \& Chelouche 2009), taking account of the redshift
dependence, to obtain $\langle \log N_{\rm HI} \rangle\approx 19.8$
(MF12), however, with a large scatter around the $W_0$ - $N_{\rm H}$
relation in mind. \textbftemp{The mean} metallicity of \MgII\ clouds is about solar
within a factor of two.  We observe that \MgII\ absorption systems \textbftemp{are}
on the \textbftemp{metallicity vs hydrogen column density relation for
DLAs}.  There seems \textbftemp{to be} no discontinuity in this relation between \MgII\ and DLA \textbftemp{absorbers}: 
the total abundances of magnesium along the column in both
absorbers are about the same.

\textbftemp{Let us now suppose that gas in \MgII\ clouds have properties similar
to typical ISM. In this case, we}
expect that
30\% of heavy elements condense into dust grains, which
  corresponds to the case when all silicate and a substantial fraction
  of graphite (15$-$50\%) condense into dust grains
(e.g., Weingartner \& Draine 2001). We also assume that the
  relation between $E(B-V)$ and $N_{\rm H}$ for the Milky Way 
holds up when scaled by metallicity (Bohlin, Savage \& Drake 1978).
\textbftemp{It then
implies that dust reddening and hydrogen column density can be used to
estimate metallicity $Z$. Numerically, this is given by the relation}
\begin{equation}
Z/Z_\odot = 
\frac
{\langle \Delta(g-i)_{\rm dust} \rangle}
{0.30\times10^{-21} k(z_{\rm abs}) N_{\rm H}}
\label{eq:metallicity}
\end{equation} 
\noindent
where 
$k(z)$ is the \lq K-correction' for the dust extinction curve for $g-i$ colour. 
There are typically two types of extinction curves, Milky Way (MW)
type and SMC type. \textbftemp{ The two extinction curves differ
little in the optical region, and eq. (\ref{eq:metallicity}) holds
for both curves up to a 20\% difference. In the UV region the two curves 
differ significantly, and it has been shown that dust in
\MgII\ clouds obeys the SMC type extinction curve (e.g., York et
al. 2006; M\'enard et al. 2008; MF12).  Which extinction curve works
for DLAs is a part of the task given to our study. 
}

\textbftemp{
The most conspicuous difference between the MW extinction
curve and the SMC curve is the presence of a hump in the former at
2175\AA.  The K-correction for MW extinction in the $g-i$ band
wildly varies as a
function of redshift for $z>0.8$ when the feature enters the $g$ band till
it goes away from the $i$ passband for $z>2.8$. The K correction
becomes even negative for $z=2.2-2.6$, which is close to the redshift
of absorbers that concern us.  This does not agree with the reddening
we observed
\footnote{We observe approximately constant reddening induced by DLAs in a
  wide range of redshift with $z\lesssim 4$.},
supporting the validity of the SMC type extinction law for DLAs. The
MW type extinction, that woud lead to very small reddening or blueing
instead, is excluded.  With the SMC type extinction curve (Weingartner \& Draine
  2001), we have $k(z_{\rm abs}=2.1)\simeq 3.0$ to $k(z_{\rm
    abs}=2.3)\simeq 3.2$. 
}

\textbfnew{
We use equation (\ref{eq:metallicity}) as a proxy to estimate metallicity.
Here we assume that the dust-to-metals ratio does not vary with the
hydrogen column density. The inferred metallicity values
can be read in the right axis of the Figure 4. 
To assess whether our mean metallicity estimate is reasonable, i.e. valid within a factor of a few,
we also plot spectroscopic metallicity estimates compiled by Neeleman et al. (2013).
From their list, we select DLAs centred around the mean redshift of our sample ($\bar z=2.2$)
and with a somewhat wider redshift range of $\pm0.3$ in order to have enough systems.
We then compute the median metallicity as a function of hydrogen column density
and show the results as green points in the figure. 
The errors are estimated by bootstrapping the sample.}
\textbfnew{It shows that our estimate of metallicity inferred from dust reddening is consistent with
spectroscopic estimates. We note that, from the spectroscopic measurements alone, 
the current data set does not allow us to obtain a statistically significant anticorrelation 
between metallicity and hydrogen column density.
This overall consistency validates the assumption that, within a factor of a few,
the median dust-to-metals ratio does not depend much on hydrogen column density in the range of values
considered in this analysis.
In addition, this agreement between the two estimates 
reassures that our procedure is not affected by substantial errors\footnote{
We here remark on the trend noted by  De Cia et al. (2013) 
that the median value of
  the dust-to-metals ratio decreases by a factor of a few when ${\rm
    N_{HI}}$ increases from about $10^{20}$ to $10^{22}\,{\rm
    cm^{-2}}$. Such an effect, if any, would have only a modest impact on the
  overall trends of the estimated metallicity as a function of column
  density seen here.}.}

The lack of DLA in the upper right plane (high $Z$, high $N_{\rm HI}$)
in Figure 4 above has been noticed in the past in spectroscopic
samples (e.g., Boiss\'e et al. 1998; P\'eroux 2003; Khare et
al. 2007).  In particular, Meiring et al. (2009) \textbftemp{ showed
  that the metallicity of DLAs is bound by ${\rm [Zn/H]}\lesssim
  -0.8\log N_{\rm HI}+{\rm constant}$.  } This lack, however, has been
ascribed to a selection effect against heavily reddened quasars, or to
sometimes more complicated atomic effects (Krumholz et al. 2009). With
only the spectroscopic sample one cannot conclude as to the cause.\\

At somewhat higher redshift we noted with a larger error that the
reddening trend does not change much from what
we observed for our $2.1<z<2.3$ sample. The 2175\AA~ hump goes off the
$i$ band for $z>2.8$, and the MW extinction curve should lead to
\textbftemp{negative} reddening. \textbftemp{The fact that we do not see a change
in the reddening of DLAs with redshift} 
from $z\approx2.1$ to higher $z$ is consistent with the shape
of the SMC extinction curve \textbftemp{but not with that of the Milky Way}.

\section{Dust in DLA and \MgII\ clouds}

We attempt to estimate the hydrogen and cosmic dust budget among
quasar line absorbing clouds. We must bear in mind that there may be
systematic effects arising from the treatment as to the selection of
the absorption system.  The numbers given below should be taken with
reservations for the errors that may not be properly estimated. We
consider, nevertheless, that it is worthwhile to envisage a global
picture as to the distribution of hydrogen and dust in intergalactic
objects.

With the SDSS DR9 catalogue Noterdaeme et al. (2012) estimated \HI\
mass density $\Omega_{\rm HI}\approx 1.0\times 10^{-3}$ at $z\simeq
2$.  We obtain with the uniform quasar sample of DR9 $\Omega_{\rm HI}
\simeq 8.6\times 10^{-4}$, using the DLA identification of Notredaeme
et al.  This value is two times larger than the HI abundance at $z=0$,
$\Omega_{\rm HI}\simeq4.2\times 10^{-4}$ (Zwaan et al. 2003), meaning
that the \HI\ mass density further decreases towards $z=0$.  We
remark, however, that Prochaska \& Wolfe (2009)'s estimate at $z\approx
2$, is lower, $4\times 10^{-4}$.  DLAs identified by Prochaska \&
Wolfe are fewer. The difference seems to arise largely from the
different treatment of the continuum around Lyman $\alpha$ in the
identification of DLAs, as well as survey path length and completeness. 
We adopt here the more recent Notredaeme et
al. catalogue without referring to this problem further.

\begin{figure}[ht]
\begin{center}
\includegraphics[width=.4\textwidth]{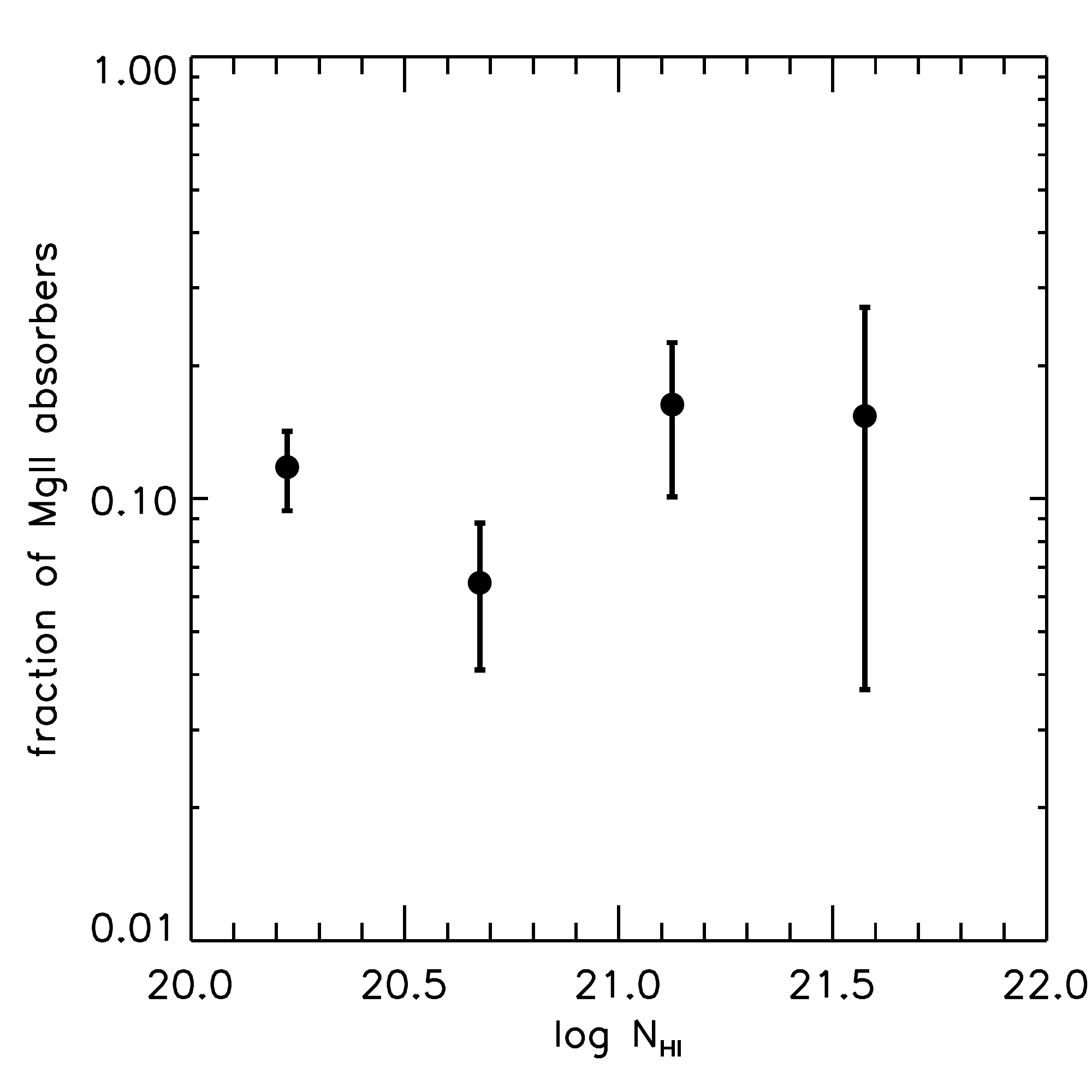}
\caption{ Fraction of clouds that show \MgII\ absorptions with
$W_0({\rm MgII})>0.8$\AA~as a function of the hydrogen column
density.
 }
\end{center}
\end{figure}

The slope of the cloud abundance as a function of the hydrogen column
density is a matter of argument (e.g., Prochaska et al. 2005; P\'eroux
et al. 2003) \textbftemp{in sub-DLA} column density, {\it i.e.} around the
Lyman limit system, LLS: $10^{17.3}<N_{\rm HI}<10^{20.3}$. Adopting
the slope $\alpha=-1.0$ of Prochaska et al. (2005, 2014) for this
column density range, we extrapolate the DLA abundance to the LLS \textbftemp{
  region} to obtain $\Omega_{\rm HI}(LLS)\approx 2-4\times 10^{-4}$,
which is half the DLA \HI\ mass density.

For our DLA sample, we estimate the incidence of \MgII\ absorption to
be approximately 10 \%.  The fraction of clouds that show
\MgII\ absorption with $W_0>0.8$ \AA~ is shown in Figure 5. It stays
roughly constant, at 10\% independent of the hydrogen column
density above $10^{20}$ cm$^{-2}$.

To account for the total incidence of \MgII\ absorbers, however, we
are led to guess that the fraction should go up to 40\% in the LLS
regime.  Adding DLAs that show \MgII\ absorption, \MgII\ clouds
amounts to $\Omega_{\rm HI}({\rm Mg II})\simeq2.1\times10^{-4}$, in
our estimate based on the \HI\ sample.  This value is not inconsistent
with our previous estimate based on the \MgII\ line selected sample,
$1.5\times10^{-4}$ (MF12), if errors from the sample selection \textbftemp{are}
considered.  Taking the fraction of \MgII\ clouds in DLAs, we estimate
the mass density of \MgII\ clouds in DLAs as $0.12\times 8.6\times
10^{-4}\simeq1\times 10^{-4}$. Subtracting this from the 
the \HI\ mass density 
in \MgII\ clouds, we estimate that the \HI\ mass density 
in \MgII\ clouds in the LLS regime is
$\approx1\times 10^{-4}$. Comparing this with $\Omega_{\rm HI}({\rm
  LLS})$ above, we infer that there may be more clouds (roughly up to
3 times more) in the LLS regime that do not show \MgII\ lines, {\it
  i.e.} are metal poor.

In MF12 we estimate that $\Omega_{\rm dust}({\rm
  MgII})\simeq2.3\times10^{-6}$.  For the DLA sample, we obtain
$\Omega_{\rm dust}({\rm DLA\& MgII})\simeq1.6\times10^{-7}$, which
should be included in the MF12 estimate.  More dust, however, is
associated with non-\MgII\ DLAs, which is estimated to be $\Omega_{\rm
  dust}({\rm DLA\& nonMgII})\simeq 5.0\times 10^{-7}$.  Dust in DLA as
a whole, $\Omega_{\rm dust}({\rm DLA)}\simeq 7\times 10^{-7}$ is still
  3 times less than that in \MgII\ clouds, namely it is only some
modest  addition to the global dust budget in absorption clouds.

We estimate dust in LLS showing \MgII\ absorption to be $2.1\times
10^{-6}$.  We are not able to estimate dust in non-\MgII\ absorbing
LLS from our samples.  It could bear an amount comparable to that in
\MgII\ clouds with the LLS hydrogen density, $\Omega_{\rm dust}({\rm
 LLS, no MgII})\sim 2\times 10^{-6}$ if we assume 1/3 the
metallicity of \MgII\ clouds, as we found for DLA.  On the other
hand, non \MgII\ LLS could be metal poor (hence dust poor) or
dominantly pristine with negligible dust, as has been uncovered by
Fumagalli et al. (2011) for their examples.  

Adding all, we estimate the amount of dust in all intergalactic absorbers $\Omega_{\rm
  dust}\approx 3-5\times 10^{-6}$.  This larger value agrees with what
we infer from reddening of rays passing in the vicinity of galaxies
(M\'enard et al. 2010, hereinafter MSFR; Fukugita 2011).  When dust in
galaxies ($4\times 10^{-6}$) (Fukugita \& Peebles 2004) is added,
there seems to be, at least, no obvious missing dust in view of the present
error, when compared with the amount that ought to be produced in the
history of galaxy (Fukugita 2011).  This consideration shows that dust
observed is consistent with the amount that is produced in stars in
galaxies ($10\times 10^{-6}$), allowing for a possibility of some
amount (say $\leq20-30$ \%) still missing, or destructed.

\begin{table}
\begin{center}
Estimates of $\Omega_{\rm HI}$ in absorbers
\begin{tabular}{c ccc}
\hline\hline
absorber & with MgII & without MgII & total\\
\hline
DLA     & $1\pm0.3\times10^{-4}$ & $8\pm2\times10^{-4}$ & $9\pm2\times10^{-4}$\\
LLS     & $\sim 1\times10^{-4}$  & $1-3\times10^{-4}$   & $2-4\times10^{-4}$\\
Ly-$\alpha$ forest & -- & -- & $0.03\times10^{-4}$\\
total   & $2\times10^{-4}$       & $10\times10^{-4}$    & $10\pm2\times10^{-4}$\\
\hline
\end{tabular}
~\\
~\\
~\\
Estimates of $\Omega_{\rm dust}$ in absorbers\\
\begin{tabular}{c ccc}
\hline\hline
absorber & with MgII & without MgII & total\\
\hline
DLA     & $0.2\times10^{-6}$ & $0.5\times10^{-6}$ & $0.7\times10^{-6}$\\
LLS     & $2.1\times10^{-6}$  & $<2\times10^{-6}$   & $2-4\times10^{-6}$\\
total   & $2.3\times10^{-6}$       & $0.5-2.5\times10^{-6}$    & $3-5\times10^{-6}$\\
\hline
\end{tabular}
\end{center}
\label{table_summary_dust}
\caption{Cosmic densities of hydrogen and dust traced by different types of absorbers. $\Omega_{\rm HI}$ estimates are based on the hydrogen column density distribution given
by Prochaska et al. (2014).}
\end{table}

%

\section{Interpretation and implications}

\textbftemp{The} low metallicity of DLAs indicates that they are primarily aggregates
of \textbftemp{ primordial} gas, rather than gas processed by stars as with \MgII\
clouds.  Metallicity of DLAs, however, is non vanishing. 
\textbftemp{ They are enriched in some way.}  \textbftemp{A} 
notable feature is that metallicity is inversely proportional to the
hydrogen column density.

We can think of two possibilities to enrich DLA with metals, star
formation activity in DLA themselves and the contamination from
outside via winds from other galaxies.  For nearby
galaxies it is established that star formation rate is proportional to
a power of hydrogen column density, known as the Schmidt (1959)-Kennicutt (1998) law,
$N_{\rm H}^{1.4}$ above the hydrogen column density threshold, $N_{\rm
  H}\approx (0.5-1)\times 10^{21}$ cm$^{-2}$. Below this column density, the
star formation efficiency drops sharply. This has also ben confirmed in a
more recent study for
low surface brightness galaxies Wyder et al. (2009).

The Schmidt-Kennicutt law leads us to expect that the column density of star
formation, hence the heavy element and dust production,
when divided by $N_{\rm H}$, is proportional to $N_{\rm H}^{+0.4}$
above some threshold. This is contrary to the trend in our analysis,
$N_{\rm H}^{-1}$.  We do not detect any threshold in the metallicity
versus hydrogen column density relation around $N_{\rm HI}\approx
(0.5-1)\times10^{21}{\rm cm}^{-2}$, which is known from Kennicutt work.  This
leads us to conclude that the observed characteristics do not agree
with the idea that dust is produced {\it in~situ} star formation in
DLA, at least, as far as the dominant part of dust is concerned.
\textbfnew{
A number of studies have characterized the level of star formation rate associated
with $z\sim2$ DLAs from continnuum emission (Rafelski et al. 2011) or emission line 
(Noterdaeme et al. 2014) measurements
and also found that the dependence on hydrogen column density does not follow the 
the Schmidt-Kennicutt law. We note that, in constrast, at low redshift
it is known that some DLAs are parts of
galaxies of a variety of morphological types (e.g., Le Brun et
al. 1997; Rao et al. 2003).
}

We argued that gas and dust in Mg II clouds are likely to be
transported by stellar activity in nearby galaxies via galactic winds
(MF12). This hypothesis may also apply to DLA. We expect that
intergalactic objects may receive deposit from intergalactic gas at
the same amount per unit area of the surface.  This means that the
average metallicity should be inversely proportional to the column
density, in agreement with what is observed here.  From the dust
distribution that extends to \textbftemp{very large scales around galaxies}
(MSFR, Fukugita 2011), we may infer that galactic winds are ubiquitous throughout
the universe. In our earlier publication (MSFR) we have shown that the
abundance ratio of dust to dark matter stays approximately constant to
a scale larger than a few Mpc away from galaxies.  Knowing that the
mean dark matter distribution obeys a power law, roughly as $r^{-2.4}$
to far beyond the virial radius of the galaxy (Masaki et al. 2012), we
infer that dust also follows a similar power law without cutoff to a
few Mpc scale.

The total global amount of dust in galaxies and absorbers is
consistent with what is ought to be produced in star
formation in galaxies. This implies that effective lifetime of dust,
including its possible regeneration, is of
the order of the age of the universe. It is somewhat puzzling to note
that a clear discontinuity is not observed between \MgII\ absorbers
and DLAs in their metallicity distributions, while the dust to gas
ratio in \MgII\ clouds implies that they are predominantly of galactic
activity origins: they should not be diluted much with pristine gas.

The amount of dust deposited to each cloud should vary: it would depend on the distance to and on dust producing activity of nearby
galaxies. We expect DLAs lying closer to galaxies to have statistically a higher metal column density.


\section{Conclusions}

\textbftemp{We have studied the dust-to-gas ratio of DLAs and used it as a 
metallicity indicator.}
The advantage is that the dust abundance can be estimated passively
from the extinction of light, provided that care is made to measure a small value of reddening.  
\textbftemp{To keep the
accuracy controlled we limited our analysis to carefully chosen redshift ranges
for both quasars and absorbers.}
Using the fact that 30\% of heavy elements
condense into dust grains, metallicity can be estimated from reddening
in broad band photometry.  This requires averaging over a large data
set and well-controlled photometry that does not suffer from arbitrary
errors.  The advantage is that the estimate of metallicity, using only
passive measurements, does not use the temperature, or does not depend
on the environment such as the radiation field.  Our central
conclusion is that average metallicity is inversely proportional to
the column density of the DLA, as $Z\sim N_{\rm HI}^{-1}$, or in other
words, the metal column density stays constant independent of the
hydrogen column density.

We argued that {\it in~situ} star formation should lead to $Z\sim
N_{\rm HI}^{+0.4}$, which contrasts to the observation of the inverse
correlation. We have not observed any threshold in the hydrogen column
density that is known for the law of star formation
(Kennicutt 1998; Wyder et al. 2009). These aspects lead us to
conclude that it is unlikely to ascribe the origin of the bulk of dust
in DLAs to {\it in~situ} star formation.  We argued \textbftemp{that, instead, }
dust in DLAs is \textbftemp{a} deposit from intergalactic space through
stellar activity \textbftemp{in the neighbourhood of} the cloud.
\textbftemp{In this case, we expect}, on average,
the same amount of dust deposited per surface area of intergalactic
clouds irrespective of the column density. It then follows that $Z\sim
N_{\rm HI}^{-1}$.  We discussed that this view does not bring a
problem into the dust budget consideration.

About 10\% of DLAs show \MgII\ absorption features at $z\sim 2$.  DLAs
that do not show strong \MgII\ absorption features  harbour 1/3
the amount of dust compared with those that show strong \MgII\
absorption. As a result, dust in DLAs as a whole resides more in those that do not
show \MgII\ absorption features. The global
amount of dust in DLAs, however, is yet 1/3 the amount in \MgII\ clouds, which
are mostly at lower, LLS column densities.

A corollary result from our study is that dust in DLA shows reddening
consistent with the SMC type extinction curve. For the redshift 
we chose ($z\approx 2.2$) the Milky Way type extinction would lead to
no colour excess or even blueing in $g-i$ colour in the presence of dust due to the
2175\AA~feature. We detected reddening when DLA is present in the
foreground, and metallicity derived using the SMC type extinction curve
agrees statistically with spectroscopic estimates.  This rules out the
Milky Way type extinction for the DLA.

DLAs seem to be aggregates of primarily unprocessed gas with small amount
of deposits from galaxy activity in their vicinity. 
\textbftemp{In constrast,} \MgII\ clouds are consistent with secondary products of galaxies,
\textbftemp{weakly} diluted with pristine gas. However, there seems to be no clear
dichotomy in the metallicity hydrogen column density relation between
the two populations. \MgII\ clouds should thoroughly be contaminated,
or even dominated by galactic wind, if they are of the primordial
origin.  There seems to be a significant population of hydrogen clouds
with hydrogen column density comparable to \MgII\ clouds but do
not show \MgII\ lines: they seem to be a population similar to DLAs
 at a lower column density extension.  
How \MgII\ clouds formed remains as an interesting problem.

\begin{acknowledgements}

  MF thanks W.M. Keck Foundation (2013) and Ambrose Monell
  Foundation (2014) at the Institute for Advanced Study, and receives
  in Tokyo Grant-in-Aid (No. 23540288) of the Ministry of Education.  BM is supported   by NSF Grant AST-1109665. He also acknowledges the hospitality of the Institute for Advanced Study.

\end{acknowledgements}

\end{document}